\documentclass[11pt]{article}
\usepackage{hyperref}
\usepackage{amssymb}
\usepackage{graphicx}
\usepackage{float}
\usepackage{slashed}
\usepackage{amsmath}
\usepackage{tabularx}
\usepackage{breqn}
\usepackage{authblk}
\usepackage{array}
\usepackage{babel}
\usepackage{verbatim}
\usepackage{dsfont}
\usepackage{cite}
\usepackage{hyperref}
\usepackage{authblk}
\usepackage{academicons}
\usepackage{xcolor}
\usepackage{physics}
\usepackage[section]{placeins}
\usepackage[a4paper, total={6.8in, 10in}]{geometry}
\numberwithin{equation}{section}
\newcommand{\be}{\begin{equation}}
\newcommand{\ee}{\end{equation}}
\newcommand{\bea}{\begin{eqnarray}}
\newcommand{\eea}{\end{eqnarray}}
\newcommand{\ba}{\begin{aligned}}
\newcommand{\ea}{\end{aligned}}
\definecolor{orcidlogocol}{HTML}{A6CE39}
\newcommand{\orcidicon}[1]{\href{https://orcid.org/#1}{\textcolor{orcidlogocol}{\aiOrcid}}}

\begin{document}

\title{High-temperature plasma in  Casimir physics}

\author[1]{Suman Kumar Panja \thanks{sumanpanja19@gmail.com}} 
\affil[1]{Centre of Excellence ENSEMBLE3 Sp. z o. o., Wolczynska Str. 133, 01-919, Warsaw, Poland}
\author[1,2]{Mathias  Bostr{\"o}m \thanks{mathias.bostrom@ensemble3.eu}}
\affil[2]{Chemical and Biological Systems Simulation Lab, Centre of New Technologies, University of Warsaw, Banacha 2C, 02-097 Warsaw, Poland}

\maketitle

\begin{abstract}
 We present a short review of an unusual but important application for a
high-temperature charged plasma. The unorthodox proposition was made by Ninham concerning a contribution from Casimir forces across high-temperature electron-positron  plasma in nuclear interactions. The key message in the current work is how high temperatures ($\sim10^{11}$ \,K) pop out as essential. Clearly, classical, semi-classical, and quantum considerations for the background media impact both the Casimir effect and the physics of stars and the Universe.
\\\\\textit{\textbf{Keywords} : Lifshitz forces, Casimir effect, Meson physics, Plasma physics } 
\end{abstract}


\section{Introduction}
\label{Intro}

In the years following Casimir's findings \,\cite{Casi,CasimirPolder48}, quantum vacuum fluctuation-induced forces have been investigated vigorously theoretically and experimentally. Vacuum and thermal fluctuations of the quantized field around a molecule or between surfaces differ from free space, resulting in attractive or repulsive intermolecular interactions. A valid concept of the nature of molecular forces was first put forward in 1894 \,\cite{Derja,Lebedev2}, ``In Hertz's researches, in his interpretation of light oscillations as electromagnetic processes, there lies another problem which has hitherto not been considered, the problem of the sources of radiation, of the processes which take place in a molecular vibrator when it radiates light energy into space.''
``This problem takes us, on the one hand, into the field of spectral analysis and, on the other, quite unexpectedly, into the theory of molecular forces, one of the most complicated problems of modern physics. This follows from the following considerations. From the standpoint of the electromagnetic theory of light, it must be admitted that between two light-emitting molecules, as between two vibrators in which electromagnetic oscillations arise, there exist mechanical forces, caused by the electrodynamic interaction of the alternating electric currents in the molecules (according to Ampere's law), or of the alternating charges in them (according to Coulomb's law). We must therefore admit that in this case there exist intermolecular forces whose origin is closely connected with radiation processes.'' Surface force measurements\,\cite{Der,TabNature,Tab,IsraTabor,White} and theoretical advancement of the Lifshitz formula \,\cite{Dzya,ParNin1969,NinhamParsegianWeiss1970,NinPars1970} of interaction energy (resulted from vacuum fluctuation of quantized field)  expanded to encompass magnetic \cite{Richmond_1971} and conductive particles, \cite{RichDavies,Davies} as well as liquids between dissimilar surfaces. In the 1970s, Ninham, and his collaborators at Australian National University initiated these developments. The development and practical implementation of the theory of intermolecular forces, and its experimental verification, even led, rumor has it, to a trio of the main contributors to being short-listed for a Nobel Prize in Chemistry. There exists in fact a considerable amount of both novel and less novel recent work done on modeling Casimir, Lifshitz, and van der Waals forces\,\cite{Lamo1997,Bost2000,Bord,Bost2001,Munday2009,SomersGarrettPalmMunday_CasimirTorque,BostromEstesoFiedlerBrevikBuhmannPerssonCarreteroParsonsCorkery2021}. A thorough discussion was given by Sernelius\,\cite{Ser2018}.
Theories of intermolecular dispersion forces have been investigated extensively that relatively little remains to be addressed \footnote{We dedicate this work to Prof. Barry W. Ninham ahead of his 90th birthday}. However, even lately, new applications arise. We will review how the temperatures relevant for stellar physics may, in fact, be in the same range as those predicted for Casimir-Yukawa forces between a pair of neutrons in an atomic nucleus.
 A recent series of papers published by Ninham et al.\,\cite{PhysRevA.67.030701,EPJDNinham2014,Ninham_Brevik_Bostrom_2022}, proposed an impact from a high-temperature plasma on the Casimir forces across an intervening electron-positron plasma for nuclear interactions. 
 Clearly, classical, semi-classical, and quantum considerations for the background media may impact the Casimir effect at the nuclear scale\,\cite{BostromAydaPalLiBrevik_Physics_2024,AydaCorkeryBrevikBostromPLA2025,PanjaAnnPhys481_2025} and in the physics of stars and the Universe.


\section{High-temperature electron-positron plasma and Casimir-Yukawa forces}
\label{Yukawa}

\subsection{Can meson physics be linked to Casimir theory?}

We next pay attention to Casimir forces between particles in the presence of a background plasma. Our interest relates to the predicted temperatures relevant for the fundamental quantum electrodynamics of nuclear interactions. Interactions among nuclear particles occur inside a plasma composed of fluctuating, continuously created, and destroyed electron-positron pairs. Nuclear particles interactions generally described by a screened Yukawa potential\,\cite{Yukawa1935}. Ninham and Bostr\"om \cite{PhysRevA.67.030701} employed an approximation to compare the screened Casimir potential with the absolute asymptotic expression of this potential across the electron-positron plasma between surfaces. Significant similarities were observed that suggested a potential contribution of screened quantum vacuum interaction between the surfaces to the interactions between the nuclear particles.\,\cite{PhysRevA.67.030701,EPJDNinham2014,Ninham_Brevik_Bostrom_2022} 
Ninham and coworkers proposed estimating two nucleons as a pair of reflecting spheres approximated by two perfectly conducting plates with nucleon cross-sectional areas. The zero temperature Casimir interaction energy and force are\,\cite{Casi},
\begin{equation}
E=-\frac{\pi^2}{720 d^3}\hbar c ~~\text{and}~~	F=-\frac{\pi^2}{240 d^4}\hbar c	
\end{equation}
Here, $d$ is the surface separation distance between two protons, $\hbar$ is the Planck constant, and $c$ the speed of light. The estimated surface area is $A=\pi\ r^2$, where $r$ is the radius of proton and taken to be \,0.8\,fermi. The surface-to-surface distance between two nucleons is estimated to be $1$ fermi. The binding energy between two nucleon, resulting from vacuum fluctuations of the quantized field, is around 5 Mev. The binding energies per nucleon typically range from 1.1 MeV to 8.8 MeV. This suggestive result was known already in the early 1970s and discussed in an unpublished manuscript by Barry Ninham and Colin Pask (recently submitted for publication in a historical journal). The problem is somewhat similar in spirit to some ideas by Casimir for the stability of charged electrons,\cite{CASIMIR1953846,BoyerPhysRev.174.1764}. Repulsive forces arise between several surface areas due to the distribution of negative charges on the surface of the electron. An attractive force has to balance this repulsive force in order to keep the electron stable and give it finite size. Casimir proposed that zero-point energy from vacuum fluctuations of quantized field may generate the attractive force known as Poincaré stress. Inspired by this idea, various calculations of Casimir energy were reported, all of which concluded that the magnitude of the interaction was right but with the incorrect sign. It gave a further repulsive force,\cite{BoyerPhysRev.174.1764}. 
Ninham and Bostr{\"o}m demonstrated almost two decades ago that screened Casimir interactions may give rise to a contribution to nuclear interactions,\cite{PhysRevA.67.030701}. Interestingly, when a nonrelativistic plasma is investigated, the relativistic energy m$c^2$ appears in the interaction energy in an interesting way: it replaces the temperature. This shows that intriguing physics might be concealed in the issue, emphasizing the importance of including a relativistic mass from the start.

\subsection{The Casimir interaction energy between perfectly reflecting surfaces across a charged plasma}
\label{NinhamParsegianWeiss}

As proposed by Lebedew\,\cite{Derja,Lebedev2}, the origin of the Casimir interaction between metal surfaces (via the fluctuations of the electromagnetic modes) is closely connected with radiation processes. 
The full formalism of quantum electrodynamics, including the formidable theory for intermolecular forces from Lifshitz and co-workers\,\cite{Dzya}, is rather complicated. The impact of the semi-classical theory\,\cite{NinhamParsegianWeiss1970}, is largely due to the fact that much of the quantum electrodynamics (QED) formalism can be obtained using Maxwell's equations by imposing proper boundary conditions when attributing to each quantized electromagnetic mode with its zero temperature ground-state energy and the free energy at finite temperature. This idea was pioneered, and explored, by Ninham and Parsegian, with co-workers and collaborators\,\cite{NinhamParsegianWeiss1970}.

For two identical planar objects in a medium, corresponding to the geometrical configuration 1|2|1, the reflection coefficients for transverse magnetic (TM) and transverse electric (TE) modes corresponding to a wave incident from medium-1 onto the interface with medium-2 are \cite{Ser2018}
\begin{equation}
r_{12,TM}
= \frac{{{\varepsilon _2}{\gamma _1} -
{\varepsilon _1}{\gamma _2}}}{{{\varepsilon _2}{\gamma _1} +
{\varepsilon _1}{\gamma _2}}},
    \label{eq:radialTM}
\end{equation}
and
\begin{equation}
r_{12,TE} = \frac{{\left( {{\gamma _1} - {\gamma _2}}
\right)}}{{\left( {{\gamma _1} + {\gamma _2}} \right)}},
    \label{eq:radialTE}
\end{equation}
respectively. In the above equation,
$\gamma_j$ is defined as
\begin{equation}
    {\gamma _i} (\omega) = \sqrt {q^2 - {\varepsilon _i}\left( \omega  \right){{\left( {\frac{\omega }{c}} \right)}^2}}, \label{eq:gamma}
\end{equation}
where ${{\varepsilon _i}\left(\omega\right)}$ is the dielectric function corresponding to the medium $i$.

The Casimir-Lifshitz between two identical planar surfaces interacting through a medium can be represented as \,\cite{Dzya,NinhamParsegianWeiss1970,Ser2018}, 
\begin{equation}
    E(d) = \hbar \sum_{\sigma=TE,TM}\int {\frac{{{d^2}q}}{{{{\left( {2\pi } \right)}^2}}}} \int\limits_0^\infty  {\frac{{d\omega }}{{2\pi }}}     \ln \left[ {{1 - e^{ - 2 \gamma _2 d} {r_{12,\sigma}}^2}\left( {i\omega } \right)} \right].
\label{eq:hamIntEnergyIntegral}
\end{equation}
Note here that the above interaction energy represents the internal energy at zero
temperature. This interaction energy at finite temperature is written as\,\cite{Dzya,NinhamParsegianWeiss1970}
\begin{equation}
    G(d,T) ={k T}\sum_{\sigma=TE,TM} \int {\frac{{d^2}q}{{{{\left( {2\pi } \right)}^2}}}}
\sum\limits_{n = 0}^\infty{\!'} {\ln \left[ {{{1 - e^{ - 2 \gamma _2 d} {r_{12,\sigma}}^2}\left( {i \xi_n }\right)}} \right]},
    \label{eq:hamIntEnergySum}
\end{equation}
where the prime denotes that the interaction energy for $n = 0$ must be halved. In Eq.~(\ref{eq:hamIntEnergyIntegral}) frequency has been replaced by discrete Matsubara frequencies as
\begin{equation}
    \xi_n = \frac{2\pi n}{\hbar \beta}; \; n = 0,\,1,\,2,\, \ldots
\end{equation}
and corresponding integral has been replaced by a summation \,\cite{Dzya,NinhamParsegianWeiss1970}.
Ninham and Pask (unpublished) observed that the Casimir interaction at zero temperature, arising from vacuum fluctuation of quantized field , is sufficient to provide the binding energy of nucleons within a nucleus. Ninham and co-workers then considered the effects of temperature,\,\cite{Dzya,NinhamParsegianWeiss1970,PhysRevA.67.030701,EPJDNinham2014,Ninham_Brevik_Bostrom_2022}

\begin{equation}
G(d,T)=\frac{k_{\mathrm B} T}{\pi}\sum_{n=0}^{\infty}{}'\int_{0}^{\infty} q\,\mathrm{d}q\;
\ln\!\left(1-e^{-2d\sqrt{q^{2}+\xi_{n}^{2}/c^{2}}}\right).
\end{equation}
Using the above, in the absence of plasma between two planner surfaces, Ninham and Daicic explicitly derived interaction energy as\,\cite{PhysRevA.57.1870,PhysRevA.67.030701,EPJDNinham2014,Ninham_Brevik_Bostrom_2022}

\begin{equation}
G(d,T)\approx \frac{- \pi^2\ \hbar c}{720d^3}- \frac{\zeta(3) k^3 T^3}{2 \pi\hbar^2 c^2}+\frac{\pi^2 d k^4 T^4}{45\hbar^3 c^3} +..,		
\end{equation}
 It is important to note that the first term in the above equation represents the attractive Casimir energy at zero temperature. The third term represents the  black body radiation energy (in vacuum and at equilibrium) between the plates. How this result contradicts the attractive Casimir interaction energy term reported in \cite{PhysRevA.67.030701}.
Following previous work we assume that the first and third terms are equal at equilibrium. From this we find a temperature in terms of the separation distance d between the plates as $T=\hbar c/2kd$, at which the attractive force balance out repulsive forces. Notably, distances of 1-2\,Fermi correspond to temperature range of $\sim10^{11}$\,K-$10^{12}$\,K. Curiously, this is in the same temperature range predicted, with an entirely different theory and entirely different underlying physics, for critical temperatures for Bose fulled star interiors. As previously identified, here the second term in the above equation, is the chemical potential term associated with the Gibbs free energy. This can be identified from existence of an electron-positron pair plasma, formed by the photon-mediated process $e^{+}+e^{-}\leftrightarrow\gamma$ \,\cite{LandauLifshitzStatPhys1} in the gap.
Using the temperature at seperation distance $d$, we can obtain the electron-positron pair density in the plasma. As discussed in \cite{LandauLifshitzStatPhys1}, the numbers of electrons and positrons in this plasma are nearly equal and both very huge, even at temperatures of the order of m$c^2$. At higher densities, the electron-positron plasma behaves more like an ideal gas, allowing ideal gas equations to be used while ignoring interparticle interactions. The second term in the above equation, can be analysed further using the given density of an electron-positron plasma,\,\cite{LandauLifshitzStatPhys1} 
\begin{equation}
\frac{\zeta(3) k^3 T^3}{2 \pi\hbar^2 c^2}=\frac{\pi \rho \hbar c}{6},	
\end{equation}
 where $\rho=\rho_-+\rho_+$. This way to express the chemical potential term leads to the equivalence we seek. 
 
 For two perfectly refelecting planes (\mbox{$r_{12}^{\mathrm{TE}}=-1$},
\mbox{$r_{12}^{\mathrm{TM}}=1$}), the vacuum fluctuation  energy given in Eq.(\ref{eq:hamIntEnergySum}) across a dissipation-free plasma takes the expression of (inserting the dielectric function for an electron-positron plasma, $\varepsilon=1+\omega_p^2/\xi_n^2$, into $\gamma_2$).
\begin{equation}
G(d,T) = \frac{{kT}}{\pi }\sum\limits_{n = 0}^\infty  {'\int_0^\infty
d } qq\ln \left[ {1 - {e^{ - 2d\sqrt {{q^2} + {{({\xi _n}/c)}^2} +
{\kappa ^2}}
}}} \right],\label{EqA11}
\end{equation}
where $\kappa=\omega_p/c$. The zero frequency term in the Matsubara sum take the alternative form\,\cite{PhysRevA.57.1870,PhysRevA.67.030701,EPJDNinham2014}
\begin{equation}
G_{n=0}(d,T) =   \frac{{kT}}{{2\pi}}  \int_\kappa^\infty dt t
\ln(1-e^{-2 d t}).
\label{EqA10}
\end{equation}
The derivation of asymptotic Casimir interaction across a plasma was a nontrivial task carried out more than 25 years ago by Ninham (for preliminary work see Ref.\,\cite{PhysRevA.57.1870,PhysRevA.67.030701}), and much later published as an appendix in the work by Ninham et al.\,\cite{EPJDNinham2014}. At fixed separation distance for temperatures sufficiently high or at fixed temperatures except for T=0\,K, for separations sufficiently large, it follows\,\cite{EPJDNinham2014} an expansion of the form
\begin{equation}
G(d,T)=-\frac{kT\kappa}{4\pi} \frac{e^{-2\kappa d}}{d} \left(1+\frac{1}{2d\kappa}\right)-\frac{(kT)^2 e^{-2 \eta d}}{\hbar c}\frac{e^{-\rho^*\eta d}}{d}+ O(e^{-4 \eta d}),
\end{equation}
where $\rho^\ast=\rho\ e^2 \hbar^2/(\pi m_e k^2\ T^2)$, $\eta=2kT/(\hbar c)$. The vacuum fluctuation energy terms for both $n=0$ and $n>0$ exhibit analogous behavior to the Yukawa potential \,\cite{EPJDNinham2014,Ninham_Brevik_Bostrom_2022}. Both contribute to the Casimir-Yukawa binding energy, remarkably aligning with the experimentally reported binding energy per nucleon \,\cite{Ninham_Brevik_Bostrom_2022}. An extension that took into account the magnetic permeability of an electron-positron was recently explored by us\,\cite{PanjaAnnPhys481_2025}.

Due to electromagnetic fluctuation interactions, the Casimir-Yukawa binding energy at this separation comprises a contribution of –0.9\, MeV coming from the $n=0$ term and –3.6\, MeV coming from the $n>0$ terms, therefore producing a total binding energy of 4.5\, MeV.  As the separation between nucleons decreases, the binding energy increases, reflecting the variation in binding energies among different nuclei. This behavior aligns with the influences of local environment on the internal structure of nucleons. The binding energy per nucleon changes among atomic nuclei, ranging from 1.1\,MeV in Deuterium to 8.8 \,MeV in Nickel-62.

\subsection{The Klein-Gordon Equation \& semi-classical estimates for the meson mass}

As pointed out by Ninham {\it{et al.}}\,\cite{Ninham_Brevik_Bostrom_2022} the vacuum fluctuation interaction energy associated with perfectly refelecting plates within a plasma can be calculated using Maxwell’s equations,\,\cite{NinhamParsegianWeiss1970,Richmond_1971} which after a Fourier transform, and exploiting the expression for the dielectric function for an electron-positron plasma, reduces to, 
\begin{equation}
\nabla^2\emptyset+\frac{\omega^2}{c^2} \left(1-\frac{\omega_p^2}{\omega^2}\right)\emptyset=0,
\label{nabla1}
\end{equation}
 Yukawa\,\cite{Yukawa1935} proposed that the interaction between nuclear particles could be obtained from the Klein-Gordon equation, which has the solution of $\emptyset_\pi\sim\pm g^2 e^{-\mu d}/d$\,\cite{Yukawa1935}. The Yukawa potential ($\emptyset_\pi$) is after a Fourier transformation given by,
\begin{equation}
\nabla^2\ \emptyset_\pi+\frac{\omega^2}{c^2}\left(1-\frac{1}{\omega^2} \left(\frac{m_\pi c^2}{\hbar}\right)^2\right) \emptyset_\pi=0.		\label{nabla2}
\end{equation}
 Wick,\,\cite{Wick} proposed that mesons operate through the emission and absorption of virtual excitations, with the time taken for the excitation to traverse between a pair of nucleons being measured from $\Delta t\sim d_\pi/c$. The relativistic energy, $\Delta E\geq m_\pi c^2$ (which adheres to the energy and time Heisenberg uncertainty principle \,\cite{LandauLifshitzQM1997}) give the relationship: $d_\pi\cong \hbar/(m_\pi c)$. From  Eq.\,\ref{nabla1} and Eq.\,\ref{nabla2} we identify  $\omega_p^2=[(m_\pi c^2)/\hbar]^2$. From this one can find meson mass as
\begin{equation}
            m_\pi=\frac{2e\hbar}{c} \sqrt{\frac{\pi\rho}{m_e c^2 }}. 		
\end{equation}     
Using the above Ninham {\it et al.}\,\cite{Ninham_Brevik_Bostrom_2022} estimated the meson mass to be $m_\pi=267m_e$ which in agreement with the experimentally reported value of (264$m_e$).

\subsection{Lifetime of  Plasmons and  Mesons}  

We assumed that, the zero-point vacuum energy and the black body radiation energy completely cancel each other at equilibrium. The remaining entities are collective excitations, specifically plasmons inside the residual electron-positron plasma. These plasmons were recognised as neutral pions. Subsequently, we estimated the lifetime of a semi-classical analogue to the $\pi_0$ meson \cite{Ninham_Brevik_Bostrom_2022}. In this lifetime, a plasmon decay into two electron-positron pairs. These can undergo decay to provide two photons. The expansion ($\Delta E$) of the plasmon peak and its duration ($\tau\geq1/\Delta E$) are both theoretically established and empirically assessed\,\cite{NinhamPhysRev.145.209},
\begin{equation}
\Delta E \sim \frac{6\pi\varepsilon_F}{5\hbar} \left(\frac{q_\pi}{q_F}\right)^2\ \left(\frac{\hbar\omega_p}{2\varepsilon_F}\right)^3\ \left[10\ln(2)+2-4.5\frac{\hbar \omega_p}{2\varepsilon_F}+O\left(\frac{\hbar \omega_p}{2\varepsilon_F}\right)^2..\right]. 
\end{equation}
In the the above equation, the Fermi energy ($\varepsilon_F\propto\rho^{2/3}$), plasma frequency ($\omega_p\propto\rho^{1/2})$, and Fermi wavevector ($q_F\propto\rho^{1/3})$. Note that all these have explicite dependence on density and in our study they have been shown to dependent on the separation distance between the nucleons. According to Ninham it is possible to without further approximations relate the wave vector with the electron and positron densities. However, after failing to find such a relation, to obtain the lifetime of the plasmon we employed an estimate inspired by Wick's arguments discussed briefly above. Specifically, we employed this reasoning to correlate the q-vector with energy. The relativistic energy associated with plasmon excitation (a meson of mass \(m_\pi\)), 
\(E \sim m_\pi c^2\)~\cite{Ninham_Brevik_Bostrom_2022}, 
is assumed to be partitioned into the kinetic energy 
\(\frac{\hbar^2 q_\pi^2}{2m_e}\) 
of each particle in two electron–positron pairs. This results in an approximation for the plasmon wave vector: $q_\pi\le c\sqrt{m_\pi\ m_e/2}/\hbar$. 
The estimate yielded an identical numerical value (to the first decimal place) as the "naive" (Weinberg's term, \cite{Weinberg2016}) Quantum Field Theory (QFT) approximation for the uncharged pion lifetime.  Both our findings for the lifetime and the "naive" estimate exhibit the same order of magnitude ($\sim0.2\times10^{-16}$s). This can be compared with QFT result ($\sim 0.80-0.852\times10^{-16}$s), which matches with the experimentally obtained result ($\sim 0.834\times10^{-16}$s).

\section{Future Outlooks}
\par We observed that advancing this field requires the extension of these concepts of nuclear interactions to incorporate a relativistic plasma response function and magnetic (spin) susceptibilities\,\cite{PanjaAnnPhys481_2025}. If the arguments supporting the contribution of vacuum fluctuation interactions to nuclear and meson physics, as  in \cite{PhysRevA.67.030701,EPJDNinham2014,Ninham_Brevik_Bostrom_2022}, which partially relate nuclear and electromagnetic interactions, are considered valid, it suggests that decomposition of nuclear forces into Coulomb and nuclear contributions may require revision. 
The issue appears to hold equal significance to that encountered in physical chemistry. The established theories are predicated on the assumption that electrostatic forces, analyzed via a nonlinear framework, and electrodynamic forces, examined using the linear approximation of Lifshitz theory, are distinct and separate. The ansatz contravenes both the Gibbs adsorption equation and the gauge requirement pertaining to the electromagnetic field.\cite{NinhamYaminsky1997} In future research, a model assumption would be that the charged $\pi^-$ and $\pi^+$ mesons emerge as bound states of electron-plasmon and positron-plasmon.

\section{Discussions: Critical Temperatures in Casimir-Yukawa}

 The aim of this mini-review is to discuss predicted critical temperatures  based on the nucleon-nucleon Casimir-Yukawa contribution to nuclear binding energy, lifetime and meson mass\,\cite{Ninham_Brevik_Bostrom_2022}.
  Remarkably, Casimir forces act between protons and neutrons on the nuclear scale\,\cite{BostromAydaPalLiBrevik_Physics_2024,AydaCorkeryBrevikBostromPLA2025,PanjaAnnPhys481_2025}.
 Notably,  the temperatures related to the creation of an electron-positron-plasma  in Casimir-Yukawa semi-classical theory for nuclear interactions is of the same order of magnitude as the estimated critical temperatures for the creation of a charged Bose-Einstein stellar core. We observe that this temperature impacts meson mass and the Casimir-Yukawa potential contribution to the nuclear binding energy. In the considered theories, it should be stressed that density, energy, and temperature are closely linked. This is similar to the beautiful observation by Wick that meson mass is related to the relevant distances via an uncertainty principle.


\vspace{6pt} 



\section*{Author Contributions}
Both authors contributed equally to the writing of this review article.

\section*{Funding}
This research is part of the project No. 2022/47/P/ST3/01236 co-funded by the National Science Centre and the European Union's Horizon 2020 research and innovation programme under the Marie Sk{\l}odowska-Curie grant agreement No. 945339. Institutional and infrastructural support for the ENSEMBLE3 Centre of Excellence was provided through the ENSEMBLE3 project (MAB/2020/14) delivered within the Foundation for Polish Science International Research Agenda Programme and co-financed by the European Regional Development Fund and the Horizon 2020 Teaming for Excellence initiative (Grant Agreement No. 857543), as well as the Ministry of Education and Science initiative “Support for Centres of Excellence in Poland under Horizon 2020” (MEiN/2023/DIR/3797).

\section*{Data availability} There were no numerical data used or generated in the current article.


\section*{Conflicts of interest} The authors declare no conflict of interest. 

\end{document}